# Nanoporous Copper Films as Platform for UV-SERS: Sensitivity and Ability to Perform Chiral Discrimination

Huaizhou Jin[a+], Anastasiia Sapunova[b+], Yanqiu Zou[c], Ali Douaki[b,d], German Lanzavecchia[b,d], Nicolo Maccaferri[e], Costantino De Angelis[f], Roman Krahne[b], Zhenrong Zheng[c], Shangzhong Jin[g*] and Denis Garoli[b,d,g*]

a. Key Laboratory of Quantum Precision Measurement, School of Physics and Optical Engineering, Zhejiang University of Technology, Hangzhou 310014, China;
b. Istituto Italiano di Tecnologia, via Morego 30, I-16163, Genova, Italy;
c. State Key Laboratory of Modern Optical Instrumentation, College of Optical Science and Engineering, Zhejiang University, Hangzhou 310027, China
d. Dipartimento di Scienze e Metodi dell'Ingegneria, Università degli Studi di Modena e Reggio Emilia, Via Amendola 2, 42122, Reggio Emilia (Italy)
e. Ultrafast Nanoscience Group, Department of Physics, Umeå University, Umeå, Sweden
f. Department of Information Engineering, University of Brescia, Via Branze, 38, Brescia 25123, Italy
g. College of Optical and Electronic Technology, China Jiliang University, Hangzhou 310018, China

*Corresponding Authors: Prof. Denis Garoli - denis.garoli@unimore.it; Prof. Shangzhong Jin - jinsz@cjlu.edu.cn
[+] equally contribution

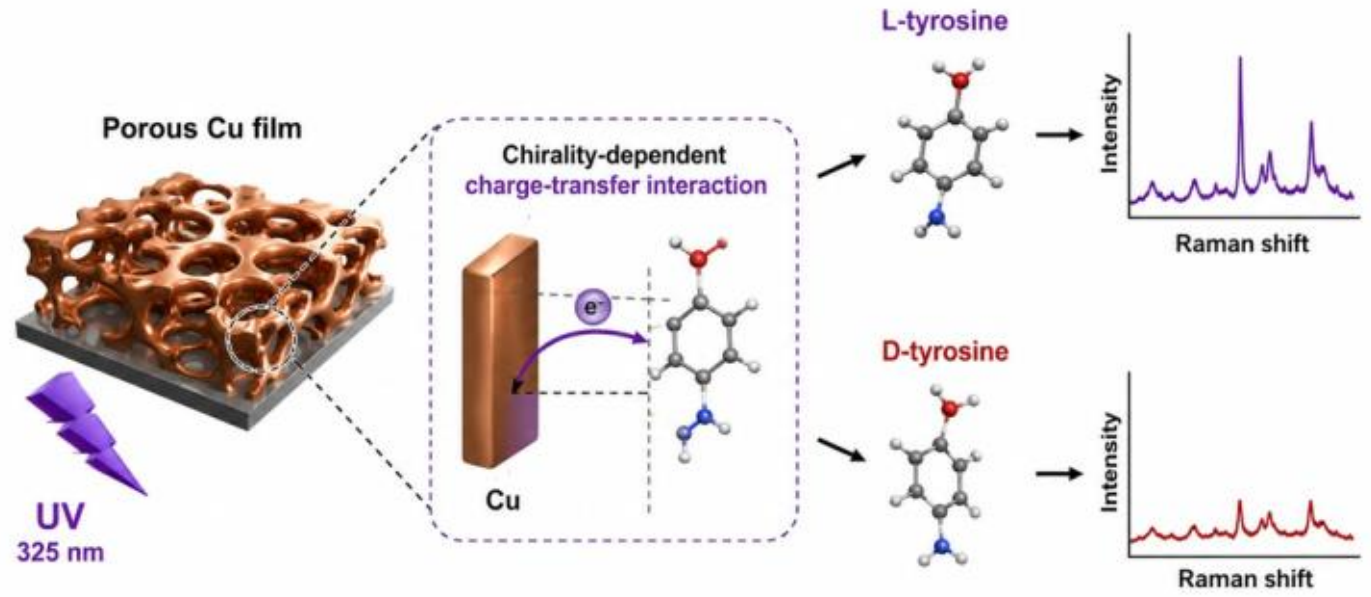

## Abstract

Surface-enhanced Raman spectroscopy (SERS) in the ultraviolet (UV) region offers important advantages for biomolecular detection, including resonance enhancement and reduced fluorescence interference. However, the development of UV-SERS substrates that combine low cost, reproducibility, and chemical stability remains challenging. Here, we employ a dry-synthesis approach to fabricate nanoporous Cu and copper oxide ($CuO_x$) films on silicon substrates and systematically evaluate their UV-SERS performance using adenine as a Raman reporter under 325 nm excitation. Among the substrates investigated, nanoporous Cu exhibits the strongest enhancement, enabling adenine detection down to 10 μM. In contrast, no detectable adenine Raman signal is observed under 532 nm excitation, indicating that the enhancement is dominated by a UV-induced chemical, charge-transfer mechanism rather than conventional electromagnetic enhancement. The Cu substrates further enable the UV Raman spectroscopy of streptavidin, as a test protein, and, more interestingly, the discrimination between L- and D-tryptophane based solely on differences in UV-SERS intensity, without chiral selectors or additional surface functionalization. By varying the substrate rotation speed during metal evaporation, the enantioselective response can be tuned, yielding L/D intensity ratios from 1.10 to 2.35 and demonstrating the critical role of substrate morphology in chiral discrimination. The dry-synthesized nanoporous films provide a simpler, scalable, and ligand-free fabrication strategy while offering additional capability for enantioselective detection. These findings establish dry-processed nanoporous Cu films as promising platforms for UV-SERS biosensing and label-free chiral analysis.

## Introduction

SERS is a highly sensitive analytical technique for trace molecular detection, relying on both electromagnetic and chemical enhancement mechanisms[1–4]. Extending SERS into the UV region is particularly interesting for biomolecular analysis, as many biologically relevant molecules (such as nucleobases, aromatic amino acids, and neurotransmitters) possess electronic transitions in the UV, enabling resonance or pre-resonance Raman enhancement[5–7]. UV excitation also reduces fluorescence background and improves spatial resolution. However, conventional gold and silver SERS substrates are unsuitable for UV operation due to strong interband damping. Alternative materials like aluminum[8–11], indium[12,13], rhodium[14–17], magnesium[18–21], and cobalt[22]have been explored, but they often present challenges related to complex fabrication, rapid oxidation, or limited enhancement.

Copper (Cu), an abundant and low-cost metal, has recently garnered attention for UV-SERS[23,24]. Although Cu exhibits weak localized surface plasmon resonance in the UV, its favorable electronic structure can support charge-transfer (CT)-mediated chemical enhancement when molecules are in direct contact. Recent studies have demonstrated UV-SERS of adenine and other aromatic molecules on nanostructured Cu surfaces. Notably, Talaikis et al. recently used femtosecond laser structuring to produce Cu films with optimal surface roughness, achieving adenine detection down to 10 μM under 325 nm excitation[23]. They attributed the enhancement to CT between Cu/$Cu_2O$ and the analyte, supported by TD-DFT calculations. Despite these advances, a comprehensive understanding of nanostructured Cu films as UV-SERS platforms remains elusive. In this work, we employ a recently established dry-synthesis method to prepare nanoporous Cu (NPCu) and $CuO_x$ films and investigate their performance in UV-SERS[25–28]. The versatility of the preparation method enables the fabrication of either metallic Cu or $CuO_x$ nanostructures simply by changing the gas used during the plasma treatment. The comparison between Cu and $CuO_x$ is motivated by recent studies reporting the potential of cuprous oxide nanostructures for UV-SERS applications[29]. Here, we show that NPCu significantly overperform $CuO_x$ structures. Our substrates exhibit sensing performance comparable to that reported in ref. [23], but, more interestingly, we additionally observe an unprecedented ability to discriminate between different enantiomers. Chiral molecules are ubiquitous in living systems, and their enantiomers often display markedly distinct physiological activities[30]. For example, tryptophan (Trp) is a fundamental amino acid for humans. For instance, L-Trp is essential for protein synthesis and serotonin metabolism, whereas D-Trp occurs in trace amounts and follows different metabolic pathways[31,32]. Hence, the development of rapid, sensitive, and label-free methods for chiral discrimination holds great significance in pharmaceutical analysis, clinical diagnostics, and food safety testing. Conventional techniques such as circular dichroism (CD) and optical rotation suffer from low sensitivity and high sample consumption, rendering them unsuitable for trace analysis[33]. On the contrary, SERS offers exceptional sensitivity, even reaching single-molecule detection[1]; however, conventional SERS substrates, whether ordered or randomly distributed metallic nanoparticles[34–39], lack intrinsic chiral selectivity. Typically, enantiomer discrimination requires modification with chiral ligands or construction of complex chiral nanostructures. For example, Liu et al.[40] synthesized chiral nanostructured gold films via amino acid induction, achieving a g-factor as high as 1.99 using the SERS-chiral anisotropy (SERS-ChA) effect; however, this approach relied on a complex symmetry-breaking growth process. Wang et al. [41] used glutathione to synthesize chiral Pt@Au triangular nanorings, again requiring chiral templates and resulting in intrinsically left- or right-handed structures. Yadav et al.[42] fabricated silver nanohelices by glancing angle deposition (GLAD), achieving chirality through helical morphology without additional ligands, yet precise control over substrate handedness was still necessary. Alternative strategies have also been explored, for instance, Bhardwaj et al. [43] employed the chirality-induced spin selectivity (CISS) effect on achiral Ag@Ni nanorod arrays combined with an external magnetic field to achieve enantiomer discrimination. Leong et al.[44] realized label-free chiral SERS detection by leveraging surface atomic defects in nanoporous gold bowls under electrochemical potential driving.

Here, we present a first proof-of-concept demonstration that NPCu films can differentiate L- and D-Trp under 325 nm excitation without any chiral functionalization. No significant L-/D-Trp difference was observed at 532 nm, whereas differentiation emerged under 325 nm excitation. This indicates the L-/D-Trp difference strongly depends on UV-excitation, and is consistent with a contribution from UV-accessible interfacial charge-transfer processes on Cu-based substrates[23,24]. Collectively, these results establish dry-processed NPCu films as a promising platform for UV-SERS-based biosensing with unexpected enantioselective sensibility.

## Results

### Comparison of UV and Visible SERS Performance of Adenine on Copper-Based Nanoporous Materials

To systematically evaluate the UV SERS sensitivity of dry-synthesized NPCu structures and compare it with that in the visible region, 325 nm and 532 nm lasers were used as excitation sources, with adenine as the Raman reporter molecule. The SERS responses of Cu and $CuO_x$ nanoporous structures were first measured probing adenine with an initial concentration 500 μM. The obtained mapping data were preprocessed and averaged; the results are shown in Figure 1. The adsorption of adenine on the copper surface may alter the electronic resonance by shifting molecular energy levels toward the laser wavelength[45,46]. Compared with non-aromatic molecules with linear-chain structures, nanostructured Cu substrates preferentially enhance the Raman bands of aromatic molecules, which is likely due to differences in their electronic structures[24]. Under 325 nm UV excitation, the NPCu substrates exhibited a clear thickness dependence (Figure 1a). Among them, the NPCu (Figure 1a, red line) prepared with 2 successive evaporations (see method section for details on the samples preparation[25,26]) showed the highest characteristic peak intensities of adenine, with well-defined peak shapes (the spectra can be compared with the control experiment performed on adenine powder reported in Supporting Information – Fig. S1) and good signal-to-noise ratio. As the thickness of the film further increased, the signal intensity decreased, which may be related to the coverage of surface active sites or changes in pore connectivity in the porous structures. In contrast, the $CuO_x$ series substrates (Figure 2b) exhibited significantly lower overall signal intensities under 325 nm excitation, with no clear thickness dependence and large data variations. Although the samples CuO-1 and CuO-4 allowed detection of some adenine characteristic peaks, their intensities were much lower than those of NPCu samples and the reproducibility of the measurements, i.e. the uniformity of the signal intensity over the sample's area, was significantly low, as illustrated by the error bars in Fig. 2b. Therefore, according to our data, it is possible to obtain some enhancement from the $CuO_x$ samples, even if electromagnetic enhancement is not expected to be significant under 325 nm excitation, given $CuO_x$ has no plasmon resonance in the UV region, and that chemical enhancement is expected to be dominant as recently discussed[29].

These results indicate that under UV excitation, NPCu exhibits superior SERS enhancement compared to $CuO_x$, and there exists an optimal thickness to be used in the NPCu film preparation. Under 532 nm visible excitation (Figure 2c-d), no distinguishable adenine characteristic peaks were detected for either Cu or $CuO_x$ substrates. This observation is consistent with the known fact that both NPCu and $CuO_x$ exhibit weak plasmonic resonance in the visible region and contributes only limited electromagnetic enhancement. Talaikis et al.[23] similarly found that adenine exhibited significant SERS enhancement under 325 nm excitation, while no SERS spectrum of adenine molecules could be detected under 532 nm excitation. They attributed this to charge transfer from Cu/$Cu_2O$ to adenine. Their TD-DFT calculations confirmed the existence of near-UV charge-transfer states between Cu and copper oxide clusters and adenine[23].

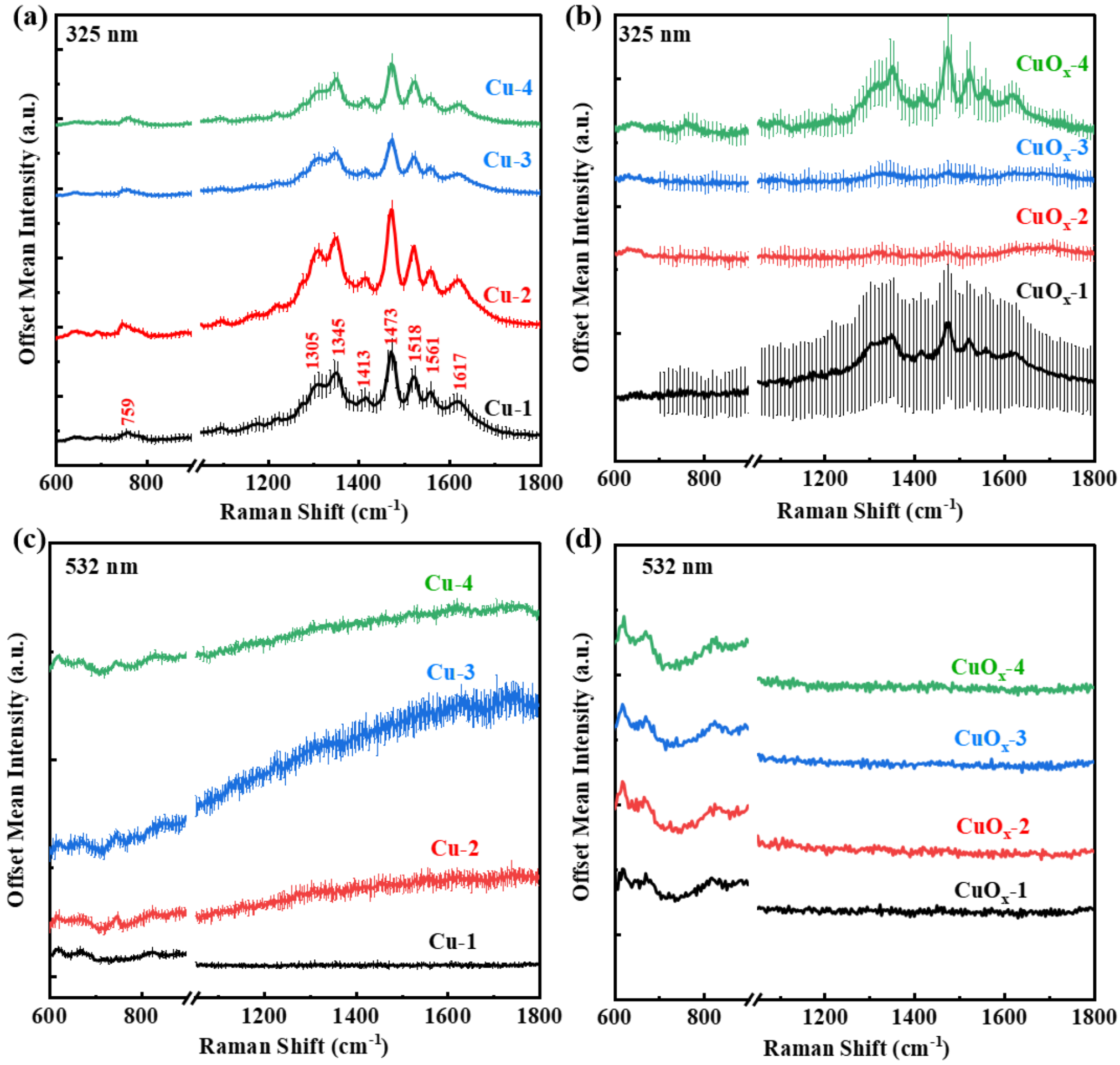

**Figure 1.** Comparison of UV/visible SERS spectra of 500 μM adenine solution on copper-based nanoporous structures. Under 325 nm excitation: (a) NPCu films with different film thickness; (b) $CuO_x$ nanoporous structures with equivalent thickness. Under 532 nm excitation: (c) NPCu structures; (d) $CuO_x$ nanoporous structures. Measurement parameters for 325 nm: 10% laser power attenuation, 5 s integration time, 1800 gr/mm grating; spectra are the average of 25 mapping points after baseline correction. Measurement parameters for 532 nm: 10% laser power attenuation, 5 s integration time, 600 gr/mm grating; spectra are the average of 25 mapping points after baseline correction

Based on comparison with the literature, the strongest bands of adenine under 325 nm excitation in this experiment can be tentatively assigned by analogy to known modes. Specifically, the strong peak at 1345 $cm^{-1}$ is attributed to the stretching motion of the C5–N7 and N1–C2 bonds (this mode appears at 1331 $cm^{-1}$ in visible SERS); the band at 1473 $cm^{-1}$ is assigned to the coupling of ν(N7–C8) with β(C8H) motion (1488 $cm^{-1}$ in visible SERS); the band at 1617 $cm^{-1}$ is related to the N9H bending mode (1601 $cm^{-1}$ in visible SERS). The band at 1561 $cm^{-1}$ can be assigned to a ring stretching mode coupled with the $NH_2$ group vibration. The observed band shifts of 10–20 $cm^{-1}$ under 325 nm excitation are consistent with the change from predominantly electromagnetic enhancement in visible SERS to mode-selective enhancement driven by chemical enhancement (CE) in the UV region. For example, the ring-breathing mode of adenine on silver surfaces has been reported at values as high as 770–783 $cm^{-1}$ [9,47]. Talaikis et al. confirmed through $H_2O/D_2O$ isotope exchange experiments that the band at ~765 $cm^{-1}$ most likely reflects the selective enhancement of adenine adsorbed on copper oxide domains; this band is sensitive to H/D exchange (shift of 9 $cm^{-1}$), further supporting its assignment to adenine adsorbed on oxygen-containing copper species[23].

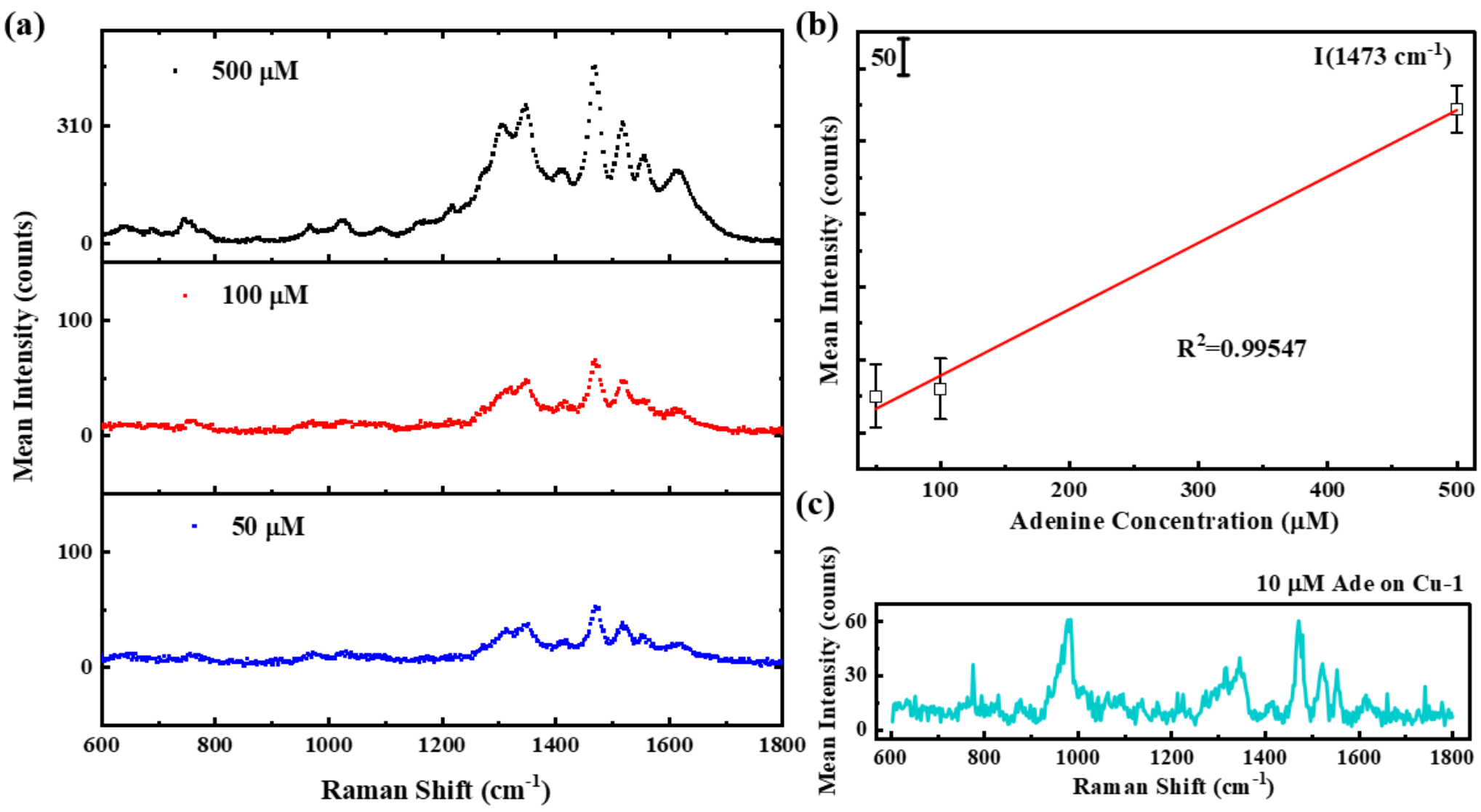

**Figure 2.** Concentration dependent UV-SERS of adenine. (a) Averaged UV-SERS spectra of adenine at concentrations ranging from 50 μM to 500 μM on a NPCu structure (Cu-2). (b) Integrated intensity of the 1472 $cm^{-1}$ band as a function of concentration, showing a near-linear response. (c) Averaged UV-SERS spectra of adenine at 10 μM concentration on NPCu structure (Cu-1). For all experiments, excitation wavelength: 325 nm, 10% laser power

attenuation, 5 s integration time, 1800 gr/mm grating; spectra are the average of 25 mapping points after baseline correction, and the error bars represent standard deviation.

To quantitatively evaluate the detection sensitivity (limit of detection) of the optimal substrate (Cu-2), we measured UV-SERS spectra of adenine at different concentrations (from 10 μM to 500 μM) under 325 nm excitation; the results are shown in Figure 2. As observed, we can clearly detect the main peaks of adenine down to a concentration of 10 μM, in agreement with the state-of-the-art for UV-SERS based on Cu nanostructures[23,24]. Surprisingly, this limit of detection is better than the one obtained using nanostructured films prepared with well-known UV plasmonic materials such as aluminum and rhodium[18,48–50]. Linear fitting of the integrated intensity of the 1472 $cm^{-1}$ band against the concentration is shown in Figure 2b. In the concentration range from 10 μM to 500 μM, the peak intensity exhibits a good linear relationship with concentration ($R^2$ = 0.9966).

**UV-SERS Response of Streptavidin**

To further verify the UV-SERS detection capability of Cu structures for biomacromolecules, we extended our experiments using streptavidin (0.05 mg/mL) as a model protein and measured the SERS response on Cu and $CuO_x$ nanoporous structures under 325 nm excitation. The experimental results are shown in Figure 3.

It can be seen that on the NPCu substrates, the characteristic Raman peaks of streptavidin are clearly detectable. A detailed analysis of the spectrum of a single streptavidin molecule under the same measurement conditions is presented in our previous report[51]. It can be observed that the Cu-2 substrate still exhibits the strongest signal intensity. The main characteristic peaks lie in the range of approximately 1000–1700 $cm^{-1}$ and can be assigned to the tryptophan-related peak (~1367 $cm^{-1}$, W7) and the ring vibration peaks of aromatic amino acids (tyrosine and tryptophan) in the region of ~1550–1620 $cm^{-1}$ [52–54], even if these peaks are partially overlapped with the peak of oxide species. As the number of layers increases to Cu-3 and Cu-4, the signal intensity gradually decreases, which is consistent with the thickness-dependent trend observed in the adenine experiments. This indicates that the optimal enhancement effect of the Cu-2 substrate is universal for both small molecules (adenine) and biomacromolecules (streptavidin).

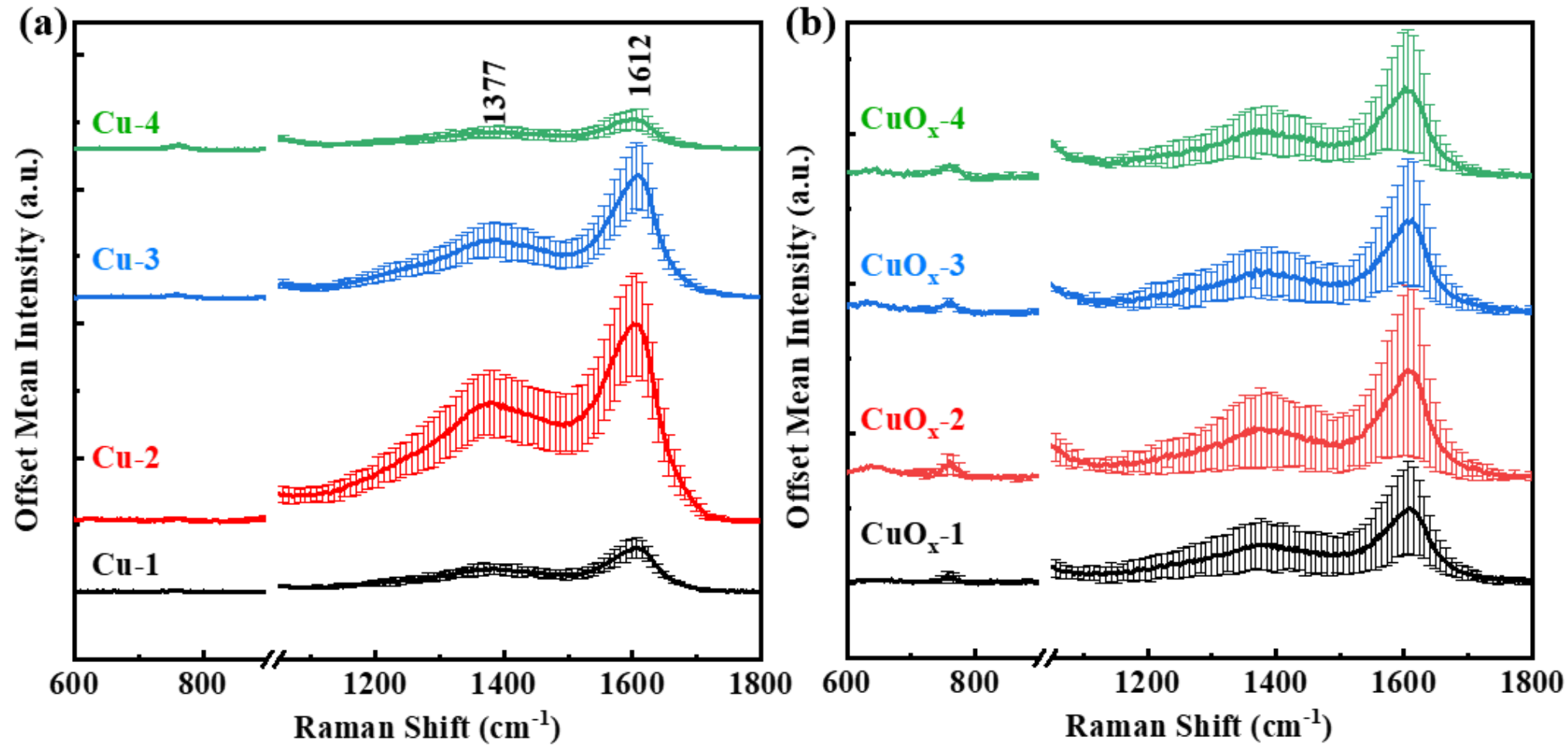

**Figure 3.** Comparison of UV-SERS spectra of 0.05 mg/mL streptavidin on copper-based nanoporous structures. (a) UV-SERS spectra on NPCu films (Cu-1 to Cu-4); (b) UV-SERS spectra on $CuO_x$ nanoporous structures ($CuO_x$-1 to $CuO_x$-4). The shaded areas represent the standard error. Excitation wavelength: 325 nm, 10% laser power attenuation, 60 s integration time, 1800 gr/mm grating; spectra are the average of three arbitrary points after baseline correction.

In contrast, on the $CuO_x$ series substrates (Figure 3b), the UV-SERS signals of streptavidin are generally weak. No significant differences in spectral peak positions or intensities are observed among different film thicknesses ($CuO_x$-1 to $CuO_x$-4), and all exhibit a low signal-to-noise ratio. Meanwhile, the standard deviations (shaded areas) of the spectra are relatively large, indicating poor signal uniformity of the $CuO_x$ substrates, which may be related to the uneven distribution of adsorption sites on the oxide surface. Although the CuO-4 substrate showed some detection capability in the adenine experiments (Figure 3b), its enhancement effect for a macromolecule such as streptavidin remains much weaker than that of the Cu-2 substrate.

**UV-SERS Analysis of Chiral Molecules Based on Copper Nanoporous Structures**

In the present work, we find that dry-synthesized NPCu structures, under UV excitation and without any chiral functionalization, can directly achieve SERS discrimination of Trp enantiomers. In particular, in our preparation method we used e-beam evaporation of Cu on a spinning substrate mounted with a tilt of 80 degrees with respect to the horizontal direction (see methods section for details). Using different rotation speeds during the evaporations of the different samples, the morphology of the final NPCu changes and can bring a chiral asymmetry. To perform proof-of-concept chiral discrimination with our NPCu we prepared three sets of samples using 1, 10 and 20 rpm, respectively.

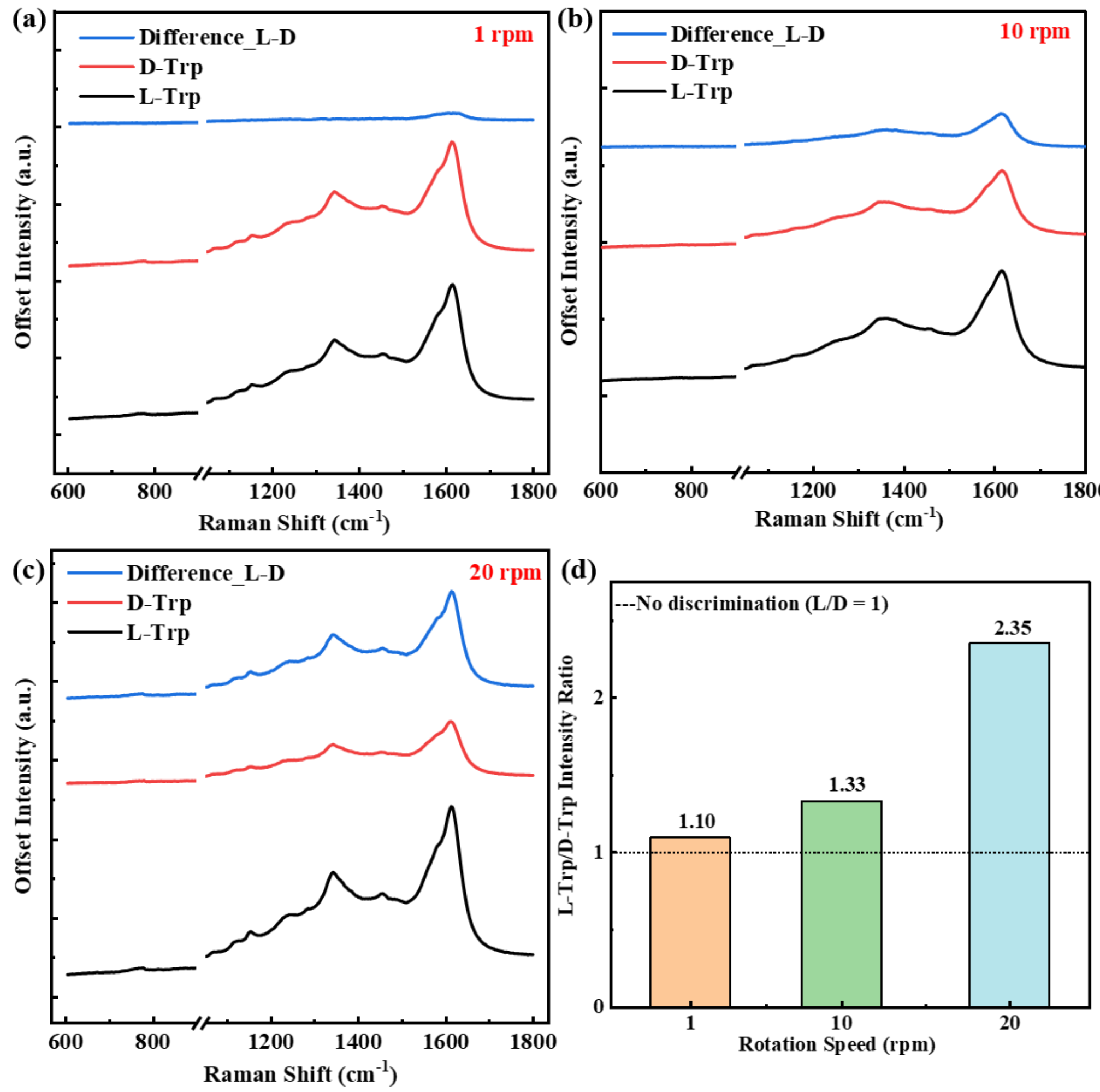

**Figure 4.** Rotation-speed dependent chiral discrimination of L- and D-tryptophan on NPCu samples. (a-c) Averaged UV-SERS spectra of 100 µM L-tryptophan (black) and D-tryptophan (red) at 20 rpm, 10 rpm, and 1 rpm rotation speeds, respectively, with blue curve being the L – D differences. (d) L- / D- Trp intensity ratio as a function of the rotation speeds used in the film preparation.

Figure 4 shows the UV-SERS spectra of 100 µM L-Trp (blue) and D-Trp (red) on NPCu substrates prepared at different rotation speeds (1, 10, and 20 rpm). The most prominent peaks are observed at ~1360 $cm^{-1}$ (W4 indole-ring C–H bending vibration) and ~1615 $cm^{-1}$ (indole aromatic-ring C=C/C–C stretching, possibly W1 indole-ring mode); the band at ~1550 $cm^{-1}$ is assigned to the W3 vibration of the aromatic ring, which is widely used as an environment-sensitive marker for the orientation of the tryptophan side chain[55,56]. The rotation speed exhibits a clear effect on chiral discrimination capability. At 1 rpm, the L-Trp and D-Trp signals are nearly equivalent (L/D intensity ratio = 1.10), indicating minimal enantioselectivity. As the rotation speed increases to 10 rpm, moderate chiral selectivity emerges (L/D = 1.33). At 20 rpm, strong chiral discrimination is achieved, with the L-Trp signal intensity 2.35× higher than that of D-Trp. The difference spectra (blue curves in Figure 4) also show that the discrimination is most pronounced at 20 rpm, with strong positive peaks at the characteristic tryptophan bands. The L-Trp/D-Trp intensity ratio was then calculated from averaged SERS spectra across 25 spatial positions for each condition. As shown in Fig. 4d, the L/D intensity ratio exhibits a clear positive correlation with rotation speed. This trend demonstrates that faster rotation enhances the substrate's ability to differentiate between enantiomers. To further support this hypothesis, we prepared an additional sample flipping its position in the evaporation chamber in order to obtain the deposition at an equivalent negative rotation speed. As reported in SI – Fig. S3, in this sample the L-Trp/D-Trp intensity ratio is inverted, showing higher intensity for the D-Trp.

The ability of achiral copper substrates to discriminate chiral molecules under UV excitation in this study can be firstly understood from the following two aspects. First, the charge transfer process between NPCu and Trp under 325 nm excitation is highly sensitive to molecular configuration. Subtle differences in the adsorption geometries of L- and D-Trp on the copper surface (e.g., the orientation of the indole ring, the coordination modes of the carboxyl/amino groups with surface copper atoms) may lead to different efficiencies of photoinduced charge transfer, thereby producing different SERS enhancement factors. Second, as proposed by Chen et al.[57], the "excitation-free enhancement" could be involved in this process. It states that in chiral plasmonic nanocavities, the chirality of the Raman signal originates from the selective enhancement of the quantum efficiency of the cavity for the chiral near field, rather than from differences in the excitation rate. Although that work was based on gold nanocube-on-mirror structures, the idea that chiral near fields can achieve enantiomer-dependent Raman signals by enhancing radiative quantum efficiency provides important inspiration for the present study. In our system, NPCu structures may generate local chiral near fields under UV excitation. Variations in surface morphology and porosity corresponding to different speeds of rotation used in the film evaporation leads to changes in the spatial distribution and intensity of the chiral near field. To better understand this potential effect, we performed Finite Elements Methods (FEM) numerical simulations using a previously reported approach based on Comsol Multiphysics. Using experimental SEM micrographs directly uploaded in the FEM tool, it is possible to calculate the electromagnetic (e.m.) field distribution in the nanostructures playing with the excitation conditions[18,25,26,58,59]. As a measure of the chirality of the electric field, we considered two different parameters: (i) the curl of the electric field, $\nabla \times E$, and (ii) the e.m. field calculated considering the nanostructures illuminated with left- and right-handed circular polarizations (here called LCP and RCP, respectively) and obtaining the $C_{enh}$=LCP-RCP. To do these calculations we considered three exemplary SEM micrographs of the samples prepared with rotation speeds of 1, 10 and 20 rpm, respectively (see SI). To keep the computation costs within an acceptable limit, we consider an area of 500x500 nm and we first calculated the e.m. field enhancement at 325 and 525 nm (see SI – Fig. S2). Successively we obtained the $|\nabla \times E|$ map (Fig. 5a) clearly showing that a rotational contribution in the e.m. field is present in all the cases. Fig. 5b reports the calculated $C_{enh}$ for the three samples, still confirming a chiral enhancement (additional details are reported in Supporting Information).

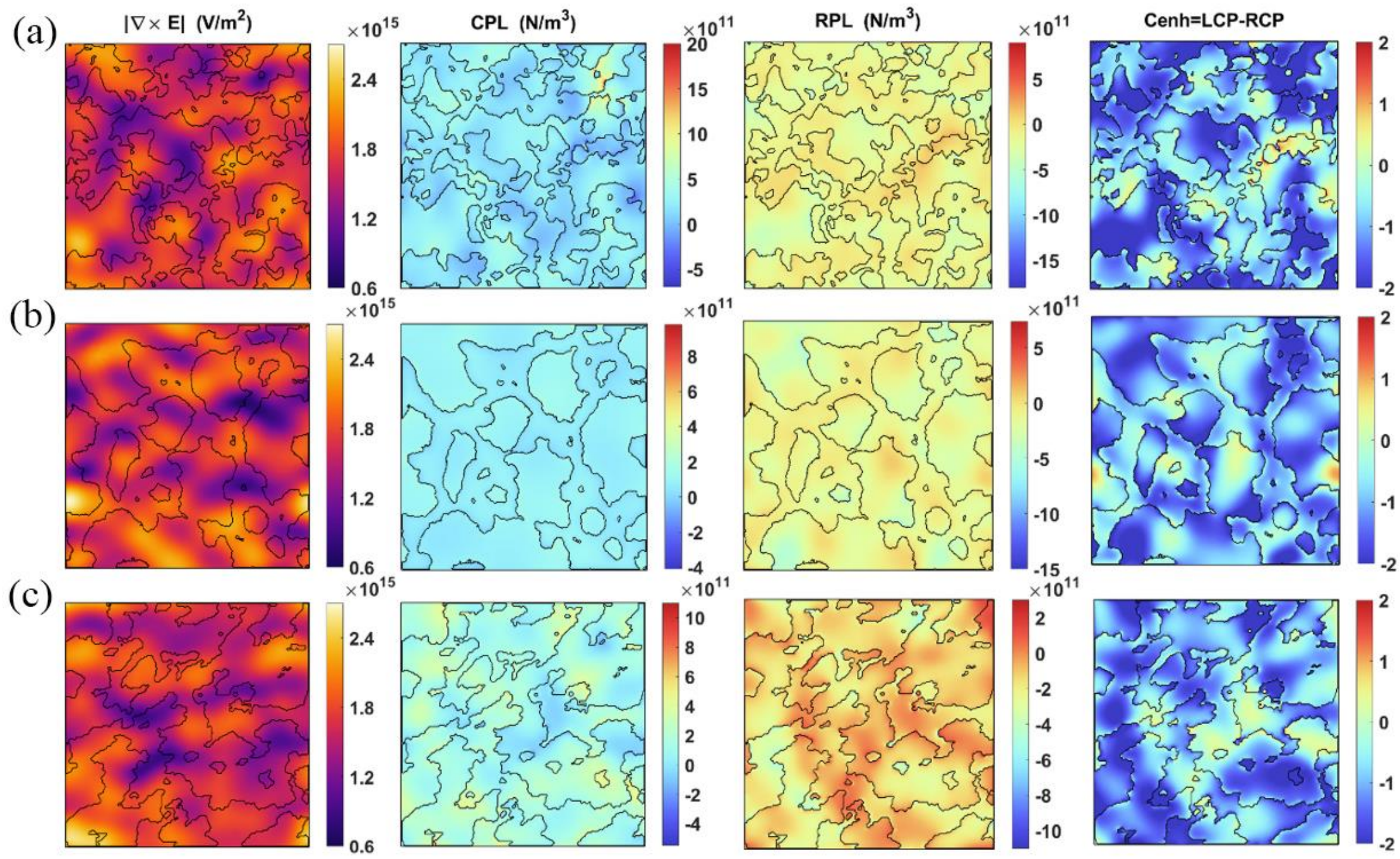

**Figure 5.** FEM simulations of the e.m. obtained considering the excitation wavelength of 325. The first column reports the $|\nabla \times E|$; the second and the third columns report the e.m. calculated considering right- and left- handed circular polarizations, respectively. The right column reports the calculated $C_{enh}$. (a) 1 rpm sample; (b) 10 rpm sample; (c) 20 rpm sample.

## Conclusions

In summary, we have demonstrated that dry-synthesized nanoporous copper films provide an effective platform for UV-SERS and enable label-free discrimination of tryptophan enantiomers. Under 325 nm excitation, NPCu exhibited substantially stronger Raman signal than the corresponding nanoporous $CuO_x$ structures, with an optimal film thickness yielding reproducible signals for both small molecules and biomolecules. Using adenine as a model analyte, characteristic Raman bands were detected down to 10 µM with a near-linear concentration response over the investigated range. In contrast, no distinguishable adenine SERS signal was observed under 532 nm excitation, supporting the predominant role of UV-induced chemical enhancement and charge-transfer processes rather than conventional electromagnetic enhancement. The applicability of the NPCu substrates to biomolecular analysis was further demonstrated by the detection of streptavidin, indicating that the detection efficiency is not limited to small molecular analytes. More importantly, NPCu enabled direct discrimination of L- and D-tryptophan without the use of chiral ligands, chiral selectors, or additional surface functionalization. The enantioselective response was strongly dependent on the substrate preparation conditions: increasing the rotation speed during oblique-angle Cu evaporation resulted in an increase of the L-Trp/D-Trp Raman intensity ratio from 1.10 at 1 rpm to 2.35 at 20 rpm. This result demonstrates that the morphology of dry-synthesized nanoporous Cu can be used to tune the chiral response of the substrate. The origin of this enantioselectivity is likely associated with the interplay between UV-induced charge-transfer enhancement and the nanoscale morphology of the NPCu surface. Differences in the adsorption configurations of the two tryptophan enantiomers may lead to different charge-transfer efficiencies, while morphology-dependent electromagnetic fields may further contribute to the observed asymmetric Raman response. The present results therefore suggest that simple nanoporous Cu substrates can exhibit enantioselective SERS responses when their nanoscale morphology and UV-excited electronic interactions are appropriately controlled. Further experimental and theoretical studies will be required to disentangle the relative contributions of chemical charge transfer, molecular adsorption geometry, and possible chiral near-field effects.

Overall, the combination of a simple, dry, and potentially scalable fabrication route with UV-SERS activity and tunable enantioselective response makes nanoporous Cu a promising platform for label-free biosensing and chiral molecular analysis. These findings broaden the potential of copper-based UV-SERS substrates beyond conventional molecular detection and highlight substrate morphology as an additional parameter for engineering selective Raman responses.

## Materials and Methods

### *Sample preparation*

The samples preparation is based on the original methods proposed in ref.[28]. In brief, poly(methyl methacrylate) (PMMA) was spin-coated on a Silicon substrate at 4,000 rpm for 2 min. Cu (>99.99 % purity) was evaporated by e-beam on a PMMA thin film at room temperature with an oblique angle of 80$^{\circ}$, a rate of 0.1 nm/s, with different rotation speeds of the sample holder (1, 10 and 20 rpm have been used) a target thickness of 13 nm was used for each evaporation. The deposited Cu film was plasma treated in $N_2$ with 200W till the whole

PMMA layer was removed. The thickness of the final film was tuned repeating the process multiple times, here Cu-1, Cu-2, Cu-3 and Cu-4 represent the samples prepared with 1, 2, 3 and 4 repeated processes, respectively. To prepare the $CuO_x$ samples, the same procedure has been followed replacing the step related to $N_2$ plasma with a plasma cleaning performed with $O_2$ with an applied power of 100W (the energy oxygen plasma produces the oxidation of the Cu nanostructures as previously demonstrated[26]). SEM micrographs of the prepared samples are reported in SI.

*Simulation*

To study the plasmonic properties of NPCu, a numerical study of the electromagnetic response of such a material was conducted using commercial software based on the finite element method (FEM) with COMSOL Multiphysics. In particular, in accordance with the procedure described in detail in our recently published works[25,26,58]the nanometer-sized pores and irregularities of nanoporous Cu were numerically reproduced using SEM images of the experimental samples. Thus, the nanoporous Cu samples obtained at 1, 10, and 20 rpm were processed through binarization and were extruded in the direction of the film thickness to reconstruct a realistic three-dimensional pore geometry reflecting the actual pore size distribution and nanoscale heterogeneities of the experimental structures. For each rotation speed, a representative periodic region measuring 500 × 500 $nm^2$ was modeled. The simulated models were illuminated at normal incidence by a plane wave with wavelengths of 325 and 532 nm, which corresponded to the excitation wavelengths used in the experiment. To assess the local field enhancement $\frac{E^2}{E_0^2}$, the ratio of the electric field strength in the presence of nanoporous metal to the amplitude of the incident field was calculated. To evaluate the chiral response of the substrates they were excited with left- and right-circularly polarized light at both wavelengths. To assess the local chirality of the near field, the curl $|\nabla \times E|$ and the differential field enhancement under left- and right-circularly polarized light (LCP and RCP, respectively), $C_{enh}$=LCP−RCP were calculated. More detailed information can be found in the SI.

*Surface Enhanced Raman Spectroscopy measurements*

SERS measurements were performed using a Horiba LabRAM HR Evolution Raman spectrometer (Horiba Jobin Yvon, Kyoto, Japan) with a 50× long-focal-length objective (NA = 0.75). Adenine with varying concentrations were measured using a 325 nm laser at 5 % power (spot diameter ∼1.03 μm) and a 532 nm laser at 5 % power (spot diameter∼0.86 μm). Spectra were acquired with a 600 grooves/mm grating, 10 s exposure time, and one accumulation. For mapping measurements, a 5 × 5 grid (total 25 points) with a 5 μm step size was performed, covering a total area of 20 × 20 $μm^2$. The same laser and acquisition parameters were maintained for each measurement point. The spectral resolution was approximately 3–4 $cm^{-1}$ and 2–3 $cm^{-1}$ for different laser wavelengths, respectively.

*Sample preparation for SERS measurements*

For each measurement, 5 µL of the corresponding analyte solution was drop-cast onto the substrate surface using a micropipette. The samples were then allowed to dry naturally at room temperature and ambient conditions before SERS measurements.

Adenine solutions with concentrations ranging from 10 to 500 µM were used for the concentration-dependent experiments. Streptavidin was measured at a concentration of 0.05 mg/mL, whereas L- and D-tryptophan were measured at a concentration of 100 µM. The exact same deposition and drying procedures were used for all analytes and substrate types.

*Sample morphological characterization*

Scanning electron microscopic (SEM) images were acquired using a Zeiss SIGMA microscope equipped with a field emission gun, operating under high vacuum at an accelerating voltage of 5 kV.

## Acknowledgments

The authors thank National Natural Science Foundation of China under Grant No. 22202167, National Natural Science Foundation of China under Grant No. 62305092, and Sichuan Science and Technology Program under Grant No. 2025YFHZ0333, the European Union under the HORIZON-Pathfinder-Open: DYNAMO, grant Agreement 101072818. Open access publishing facilitated by Istituto Italiano di Tecnologia, as part of the Wiley - CRUI- CARE agreement.

## References

(1) Zou, Y.; Jin, H.; Ma, Q.; Zheng, Z.; Weng, S.; Kolataj, K.; Acuna, G.; Bald, I.; Garoli, D. Advances and Applications of Dynamic Surface-Enhanced Raman Spectroscopy (SERS) for Single Molecule Studies. *Nanoscale* **2025**, *7*, 3656–3670. https://doi.org/10.1039/D4NR04239E.

(2) Jin, H.; Cai, Y.; Song, C.; Jin, S.; Lin, Q. Advances in Single-Molecule Surface-Enhanced Raman Spectroscopy (SERS) for Biosensing. *Vib. Spectrosc.* **2025**, *138*, 103784. https://doi.org/10.1016/j.vibspec.2025.103784.

(3) Yi, J.; You, E.; Hu, R.; Wu, D.; Liu, G.; Yang, Z.-L.; Zhang, H.; Gu, Y.; Wang, Y.-H.; Wang, X.; Ma, H.; Yang, Y.; Liu, J.-Y.; Fan, F. R.; Zhan, C.; Tian, J.; Qiao, Y.; Wang, H.; Luo, S.-H.; Meng, Z.; Mao, B.-W.; Li, J.-F.; Ren, B.; Aizpurua, J.; Apkarian, V. A.; Bartlett, P. N.; Baumberg, J.; Bell, S. E. J.; Brolo, A. G.; Brus, L. E.; Choo, J.; Cui, L.; Deckert, V.; Domke, K. F.; Dong, Z.; Duan, S.; Faulds, K.; Frontiera, R.; Halas, N.; Haynes, C.; Itoh, T.; Kneipp, J.; Kneipp, K.; Ru, E. C. L.; Li, Z.-P.; Ling, X. Y.; Lipkowski, J.; Liz-Marzán, L. M.; Nam, J.-M.; Nie, S.; Nordlander, P.; Ozaki, Y.; Panneerselvam, R.; Popp, J.; Russell, A. E.; Schlücker, S.; Tian, Y.; Tong, L.; Xu, H.; Xu, Y.; Yang, L.; Yao, J.; Zhang, J.; Zhang, Y.; Zhang, Y.; Zhao, B.; Zenobi, R.; Schatz, G.

C.; Graham, D.; Tian, Z.-Q. Surface-Enhanced Raman Spectroscopy: A Half-Century Historical Perspective. *Chem. Soc. Rev.* **2025**, *54* (3), 1453–1551. https://doi.org/10.1039/D4CS00883A.

(4) Lee, Y.; Choi, K.; Kim, J.; Cha, S.; Nam, J. Integrating, Validating, and Expanding Information Space in Single-Molecule Surface-Enhanced Raman Spectroscopy for Biomolecules. *ACS Nano* **2024**, acsnano.4c09218. https://doi.org/10.1021/acsnano.4c09218.

(5) Zhao, D.; Lin, Z.; Zhu, W.; Lezec, H.; Xu, T.; Agrawal, A.; Zhang, C.; Huang, K. Recent Advances in Ultraviolet Nanophotonics: From Plasmonics and Metamaterials to Metasurfaces. *NIST* **2021**, *10* (9).

(6) Kumamoto, Y.; Taguchi, A.; Kawata, S. Deep-Ultraviolet Biomolecular Imaging and Analysis. *Adv. Opt. Mater.* **2019**, *7* (5), 1801099. https://doi.org/10.1002/adom.201801099.

(7) Hendry, E.; Carpy, T.; Johnston, J.; Popland, M.; Mikhaylovskiy, R. V.; Lapthorn, A. J.; Kelly, S. M.; Barron, L. D.; Gadegaard, N.; Kadodwala, M. Ultrasensitive Detection and Characterization of Biomolecules Using Superchiral Fields. *Nat. Nanotechnol.* **2010**, *5* (11), 783–787. https://doi.org/10.1038/nnano.2010.209.

(8) Dubey, A.; Mishra, R.; Cheng, C.; Kuang, Y.; Gwo, S.; Yen, T. Demonstration of a Superior Deep-UV Surface-Enhanced Resonance Raman Scattering (SERRS) Substrate and Single-Base Mutation Detection in Oligonucleotides. *J. Am. Chem. Soc.* **2021**, *143* (46), 19282–19286. https://doi.org/10.1021/jacs.1c09762.

(9) Sharma, B.; Cardinal, M. F.; Ross, M. B.; Zrimsek, A. B.; Bykov, S. V.; Punihaole, D.; Asher, S. A.; Schatz, G. C.; Van Duyne, R. P. Aluminum Film-over-Nanosphere Substrates for Deep-UV Surface-Enhanced Resonance Raman Spectroscopy. *Nano Lett.* **2016**, *16* (12), 7968–7973. https://doi.org/10.1021/acs.nanolett.6b04296.

(10) Jha, S. K.; Ahmed, Z.; Agio, M.; Ekinci, Y.; Löffler, J. F. Deep-UV Surface-Enhanced Resonance Raman Scattering of Adenine on Aluminum Nanoparticle Arrays. *J. Am. Chem. Soc.* **2012**, *134* (4), 1966–1969. https://doi.org/10.1021/ja210446w.

(11) Ding, T.; Sigle, D. O.; Herrmann, L. O.; Wolverson, D.; Baumberg, J. J. Nanoimprint Lithography of Al Nanovoids for Deep-UV SERS. *ACS Appl. Mater. Interfaces* **2014**, *6* (20), 17358–17363. https://doi.org/10.1021/am505511v.

(12) Kumamoto, Y.; Taguchi, A.; Honda, M.; Watanabe, K.; Saito, Y.; Kawata, S. Indium for Deep-Ultraviolet Surface-Enhanced Resonance Raman Scattering. *ACS Photonics* **2014**, *1* (7), 598–603. https://doi.org/10.1021/ph500076k.

(13) Das, R.; Soni, R. K. Highly Stable In@SiO2 Core-Shell Nanostructures for Ultraviolet Surface-Enhanced Raman Spectroscopy. *Appl. Surf. Sci.* **2019**, *489*, 755–765. https://doi.org/10.1016/j.apsusc.2019.06.003.

(14) Watson, A. M.; Zhang, X.; Alcaraz De La Osa, R.; Sanz, J. M.; González, F.; Moreno, F.; Finkelstein, G.; Liu, J.; Everitt, H. O. Rhodium Nanoparticles for Ultraviolet Plasmonics. *Nano Lett.* **2015**, *15* (2), 1095–1100. https://doi.org/10.1021/nl5040623.

(15) Kumar, G.; Soni, R. K. Rhodium Concave Nanocubes and Nanoplates as Deep-UV Resonant SERS Platform. *J. Raman Spectrosc.* **2022**, *53* (11), 1890–1903. https://doi.org/10.1002/jrs.6427.

(16) Muñeton Arboleda, D.; Coviello, V.; Palumbo, A.; Pilot, R.; Amendola, V. Rhodium Nanospheres for Ultraviolet and Visible Plasmonics. *Nanoscale Horiz.* **2025**, *10* (2), 336–348. https://doi.org/10.1039/D4NH00449C.

(17) Xu, Y. Rounding up Rh Nanoparticles for Ultraviolet Plasmonic Sensing. *Nanoscale Horiz.* **2025**, *10* (4), 659–661. https://doi.org/10.1039/D5NH90005K.

(18) Ponzellini, P.; Giovannini, G.; Cattarin, S.; Zaccaria, R. P.; Marras, S.; Prato, M.; Schirato, A.; D'Amico, F.; Calandrini, E.; De Angelis, F.; Yang, W.; Jin, H.-J.; Alabastri, A.; Garoli, D. Metallic Nanoporous Aluminum–Magnesium Alloy for UV-Enhanced Spectroscopy. *J. Phys. Chem. C* **2019**, *123* (33), 20287–20296. https://doi.org/10.1021/acs.jpcc.9b04230.

(19) Gutiérrez, Y.; Alcaraz De La Osa, R.; Ortiz, D.; Saiz, J.; González, F.; Moreno, F. Plasmonics in the Ultraviolet with Aluminum, Gallium, Magnesium and Rhodium. *Appl. Sci.* **2018**, *8* (1), 64–77. https://doi.org/10.3390/app8010064.

(20) Honda, M.; Hizumi, K.; Kataoka, A. Nanostructure of Al–Mg System as Novel UV Plasmonic Material. *Opt. Commun.* **2025**, *591*, 132131. https://doi.org/10.1016/j.optcom.2025.132131.

(21) Giordano, A. N.; Rao, R. Beyond the Visible: A Review of Ultraviolet Surface-Enhanced Raman Scattering Substrate Compositions, Morphologies, and Performance. *Nanomaterials* **2023**, *13* (15), 2177. https://doi.org/10.3390/nano13152177.

(22) Remeikienė, A.; Matulaitienė, I.; Selskis, A.; Talaikis, M.; Niaura, G. Electrochemical UV-SERS of Adenine on Cobalt Electrode. *Spectrochim. Acta. A. Mol. Biomol. Spectrosc.* **2025**, *330*, 125733. https://doi.org/10.1016/j.saa.2025.125733.

(23) Talaikis, M.; Liudvinavičius, R.; Stankevičius, E.; Selskienė, A.; Gkouzi, A.-M.; Murauskas, T.; Sivakov, V.; Niaura, G. Femtosecond Laser-Induced Nanostructures in Copper Film for UV-SERS. *ACS Appl. Mater. Interfaces* **2025**, acsami.5c21582. https://doi.org/10.1021/acsami.5c21582.

(24) Yadav, S.; Talaikis, M.; Ryabchikov, Y. V.; Ziegler, M.; Niaura, G.; Deckert-Gaudig, T.; Sivakov, V. Copper-Based Multiwavelength UV Surface Enhanced Raman Spectroscopy. *Adv. Opt. Mater.* **2025**, *13* (18), 2500078. https://doi.org/10.1002/adom.202500078.

(25) Caligiuri, V.; Kwon, H.; Griesi, A.; Ivanov, Y. P.; Schirato, A.; Alabastri, A.; Cuscunà, M.; Balestra, G.; De Luca, A.; Tapani, T.; Lin, H.; Maccaferri, N.; Krahne, R.; Divitini, G.; Fischer, P.; Garoli, D. Dry Synthesis of Bi-Layer Nanoporous Metal Films as Plasmonic Metamaterial. *Nanophotonics* **2024**, *13* (7), 1159–1167. https://doi.org/10.1515/nanoph-2023-0942.

(26) Tapani, T.; Caligiuri, V.; Zou, Y.; Griesi, A.; Ivanov, Y. P.; Cuscunà, M.; Balestra, G.; Lin, H.; Sapunova, A.; Franceschini, P.; Tognazzi, A.; De Angelis, C.; Divitini, G.; Carzino, R.; Kwon, H.; Fischer, P.; Krahne, R.; Maccaferri, N.; Garoli, D. Disordered Plasmonic System with Dense Copper Nano-Island Morphology. *Nanophotonics* **2025**. https://doi.org/10.1515/nanoph-2024-0743.

(27) Tapani, T.; Pettersson, J. M.; Henriksson, N.; Brunner, C. M.; Zimmermann, A. C.; Zäll, E.; Hauff, N. V.; Das, L.; Sapunova, A.; Balestra, G.; Cuscunà, M.; De Andrés, A.; Giovannini, T.; Garoli, D.; Maccaferri, N. Morphology-Modified Contributions of Electronic Transitions to the Optical Response of Plasmonic Nanoporous Gold Metamaterial. *Nat. Commun.* **2026**, *17* (1), 829–839. https://doi.org/10.1038/s41467-026-68506-0.

(28) Kwon, H.; Barad, H.; Silva Olaya, A. R.; AlarcónCorrea, M.; Hahn, K.; Richter, G.; Wittstock, G.; Fischer, P. Dry Synthesis of Pure and Ultrathin Nanoporous Metallic Films. *ACS Appl. Mater. Interfaces* **2023**, *15* (4), 5620–5627. https://doi.org/10.1021/acsami.2c19584.

(29) Talaikis, M.; Petrulevičienė, M.; Čepėnas, R.; Pudžaitis, V.; Savickaja, I.; Pakštas, V.; Naujokaitis, A.; Mikoliūnaitė, L.; Šablinskas, V.; Niaura, G. UV-SERRS and SEIRAS Study of Adenine Adsorption on Cuprous Oxide Nanostructures. *Spectrochim. Acta. A. Mol. Biomol. Spectrosc.* **2026**, *363*, 128346. https://doi.org/10.1016/j.saa.2026.128346.

(30) Kong, X.-T.; Besteiro, L. V.; Wang, Z.; Govorov, A. O. Plasmonic Chirality and Circular Dichroism in Bioassembled and Nonbiological Systems: Theoretical Background and Recent

Progress. *Adv. Mater.* **2020**, *32* (41), 1801790. https://doi.org/10.1002/adma.201801790.

(31) Yin, J.; Zhang, B.; Yu, Z.; Hu, Y.; Lv, H.; Ji, X.; Wang, J.; Peng, B.; Wang, S. Ameliorative Effect of Dietary Tryptophan on Neurodegeneration and Inflammation in D -Galactose-Induced Aging Mice with the Potential Mechanism Relying on AMPK/SIRT1/PGC-1α Pathway and Gut Microbiota. *J. Agric. Food Chem.* **2021**, *69* (16), 4732–4744. https://doi.org/10.1021/acs.jafc.1c00706.

(32) Wang, F.; Du, R.; Shang, Y. Biological Function of D-Tryptophan: A Bibliometric Analysis and Review. *Front. Microbiol.* **2025**, *15*, 1455540. https://doi.org/10.3389/fmicb.2024.1455540.

(33) Zukowski, J.; Tang, Y.; Berthod, A.; Armstrong, D. W. Investigation of a Circular Dichroism Spectrophotometer as a Liquid Chromatography Detector for Enantiomers: Sensitivity, Advantages and Limitations. *Anal. Chim. Acta* **1992**, *258* (1), 83–92. https://doi.org/10.1016/0003-2670(92)85200-P.

(34) Tian, Y.; Fang, G.; Wu, F.; Kauno, J. G.; Wei, H.; Hsu, H.-Y.; Li, F.; Xu, G.; Niu, W. Raman Spectroscopic Technologies for Chiral Discrimination: Current Status and New Frontiers. *Coord. Chem. Rev.* **2025**, *526*, 216375. https://doi.org/10.1016/j.ccr.2024.216375.

(35) Arabi, M.; Ostovan, A.; Wang, Y.; Mei, R.; Fu, L.; Li, J.; Wang, X.; Chen, L. Chiral Molecular Imprinting-Based SERS Detection Strategy for Absolute Enantiomeric Discrimination. *Nat. Commun.* **2022**, *13* (1), 5757. https://doi.org/10.1038/s41467-022-33448-w.

(36) Ma, F.; Dou, H.; Luo, D.; Yan, Y.; Zhou, J.; Xu, G.; Wang, Y.; Zhao, L. Recent Advances in Chiral Recognition Based on Surface-Enhanced Raman Scattering Spectroscopy. *Chin. Chem. Lett.* **2025**, 111632.

(37) Abbas, A.; Zhang, Q.; Kazmi, J.; Li, Y.; Li, W.; Ahmad, W.; Zou, C.; Liang, Q. Label-Free SERS Fingerprinting for Chiral Discrimination Using Chiral Two-Dimensional Superlattice. *Nano Lett.* **2025**, *25* (33), 12676–12684. https://doi.org/10.1021/acs.nanolett.5c03047.

(38) Wu, Y.; Yang, Y.; Wang, R.; Yang, H.; Liu, X. Multiscale Synergistic SERS Chiral Detection: Gold Nanoclusters-Mediated Enantioselective Molecular Discrimination on the Self-Assembly of Gold Nanoparticles. *Anal. Chem.* **2025**, *97* (31), 17085–17093. https://doi.org/10.1021/acs.analchem.5c02859.

(39) Cong, W.; Du, Y.; Li, H.; Li, H.; Niu, W.; Wang, S.; Wong, K.-Y.; Zheng, G. Chiral Imprinting on Discrete Homochiral Au Nanohelicoids for Free Chiral Analyst Recognition with High Enantioselectivity. *J. Colloid Interface Sci.* **2025**, 139513.

(40) Liu, Z.; Ai, J.; Kumar, P.; You, E.; Zhou, X.; Liu, X.; Tian, Z.; Bouř, P.; Duan, Y.; Han, L.; Kotov, N. A.; Ding, S.; Che, S. Enantiomeric Discrimination by Surface-Enhanced Raman Scattering–Chiral Anisotropy of Chiral Nanostructured Gold Films. *Angew. Chem. Int. Ed.* **2020**, *59* (35), 15226–15231. https://doi.org/10.1002/anie.202006486.

(41) Wang, G.; Hao, C.; Ma, W.; Qu, A.; Chen, C.; Xu, J.; Xu, C.; Kuang, H.; Xu, L. Chiral Plasmonic Triangular Nanorings with SERS Activity for Ultrasensitive Detection of Amyloid Proteins in Alzheimer's Disease. *Adv. Mater.* **2021**, *33* (38), 2102337. https://doi.org/10.1002/adma.202102337.

(42) Yadav, S.; Bhardwaj, L.; Singh, J. P. Ultrasensitive Plasmonic Chiral SERS Substrate for Ligand-Free Enantioselective Discrimination and Molecular Fingerprinting. *ACS Appl. Mater. Interfaces* **2025**, *17* (32), 46383–46395. https://doi.org/10.1021/acsami.5c11104.

(43) Bhardwaj, L.; Yadav, J.; Yadav, S.; Singh, J. P. Spin-Selective Charge Transfer-SERS Driven Label-Free Enantioselective Discrimination of Chiral Molecules on Ag Nanoparticle-Decorated

Ni Nanorod Arrays. *ACS Appl. Mater. Interfaces* **2024**, *16* (49), 67289–67301. https://doi.org/10.1021/acsami.4c14701.

(44) Leong, S. X.; Koh, C. S. L.; Sim, H. Y. F.; Lee, Y. H.; Han, X.; Phan-Quang, G. C.; Ling, X. Y. Enantiospecific Molecular Fingerprinting Using Potential-Modulated Surface-Enhanced Raman Scattering to Achieve Label-Free Chiral Differentiation. *ACS Nano* **2021**, *15* (1), 1817–1825. https://doi.org/10.1021/acsnano.0c09670.

(45) Chaudhry, I.; Hu, G.; Ye, H.; Jensen, L. Toward Modeling the Complexity of the Chemical Mechanism in SERS. *ACS Nano* **2024**, *18* (32), 20835–20850. https://doi.org/10.1021/acsnano.4c07198.

(46) Valley, N.; Greeneltch, N.; Van Duyne, R. P.; Schatz, G. C. A Look at the Origin and Magnitude of the Chemical Contribution to the Enhancement Mechanism of Surface-Enhanced Raman Spectroscopy (SERS): Theory and Experiment. *J. Phys. Chem. Lett.* **2013**, *4* (16), 2599–2604. https://doi.org/10.1021/jz4012383.

(47) Kämmer, E.; Olschewski, K.; Bocklitz, T.; Rösch, P.; Weber, K.; Cialla, D.; Popp, J. A New Calibration Concept for a Reproducible Quantitative Detection Based on SERS Measurements in a Microfluidic Device Demonstrated on the Model Analyte Adenine. *Phys. Chem. Chem. Phys.* **2014**, *16* (19), 9056. https://doi.org/10.1039/c3cp55312d.

(48) Zou, Y.; Mattarozzi, L.; Jin, H.; Ma, Q.; Cattarin, S.; Weng, S.; Douaki, A.; Lanzavecchia, G.; Kołątaj, K.; Corduri, N.; Johns, B.; Maccaferri, N.; Acuna, G.; Zheng, Z.; Jin, S.; Garoli, D. UV-SERS Monitoring of Plasmonic Photodegradation of Biomolecules on Aluminum Platforms Decorated with Rhodium Nanoparticles. *Nanoscale Adv.* **2025**, *7* (17), 5212–5220. https://doi.org/10.1039/D5NA00486A.

(49) Banerjee, S.; Mattarozzi, L.; Maccaferri, N.; Cattarin, S.; Weng, S.; Douaki, A.; Lanzavecchia, G.; Sapunova, A.; D'Amico, F.; Ma, Q.; Zou, Y.; Krahne, R.; Kneipp, J.; Garoli, D. Porous Aluminum Decorated with Rhodium Nanoparticles: Preparation and Use as a Platform for UV SERS. *Mater. Adv.* **2024**, *5*, 6248–6254. https://doi.org/10.1039/D4MA00203B.

(50) Garoli, D.; Schirato, A.; Giovannini, G.; Cattarin, S.; Ponzellini, P.; Calandrini, E.; Proietti Zaccaria, R.; D'Amico, F.; Pachetti, M.; Yang, W.; Jin, H.-J.; Krahne, R.; Alabastri, A. Galvanic Replacement Reaction as a Route to Prepare Nanoporous Aluminum for UV Plasmonics. *Nanomaterials* **2020**, *10* (1), 102. https://doi.org/10.3390/nano10010102.

(51) Zou, Y.; Corduri, N.; D'Amico, F.; Kołątaj, K.; Jin, H.; Zheng, Z.; Yu, Y.; Liu, J.; Weng, S.; Douaki, A.; Krahne, R.; Wenger, J.; Jin, S.; Acuna, G.; Garoli, D. Self-Assembled Rhodium Nanoantennas for Single-Protein UV SERS. *Nanophotonics* **2026**, *15* (10), e70137. https://doi.org/10.1002/nap2.70137.

(52) Dass, M.; Gür, F. N.; Kołątaj, K.; Urban, M. J.; Liedl, T. DNA Origami-Enabled Plasmonic Sensing. *J. Phys. Chem. C* **2021**, *125* (11), 5969–5981. https://doi.org/10.1021/acs.jpcc.0c11238.

(53) Clarkson, J.; Batchelder, D. N.; Smith, D. A. UV Resonance Raman Study of Streptavidin Binding of Biotin and 2-Iminobiotin: Comparison with Avidin. *Biopolymers* **2001**, *62* (6), 307–314. https://doi.org/10.1002/bip.10003.

(54) Schuknecht, F.; Kołątaj, K.; Steinberger, M.; Liedl, T.; Lohmueller, T. Accessible Hotspots for Single-Protein SERS in DNA-Origami Assembled Gold Nanorod Dimers with Tip-to-Tip Alignment. *Nat. Commun.* **2023**, *14* (1), 7192–7202. https://doi.org/10.1038/s41467-023-42943-7.

(56) Jacob, C. R.; Luber, S.; Reiher, M. Calculated Raman Optical Activity Signatures of Tryptophan Side Chains. *ChemPhysChem* **2008**, *9* (15), 2177–2180. https://doi.org/10.1002/cphc.200800448.

(57) Chen, Y.; Wu, H.; Zhang, K.; Shao, L.; Wang, J. Quantifying the Selective Excitation-Free Enhancement of Chiral Plasmonic Nanocavities by Raman Scattering. *Nano Lett.* **2026**, *26* (18), 6112–6117. https://doi.org/10.1021/acs.nanolett.6c00805.

(58) Zou, Y.; Sapunova, A.; Giovannini, T.; Wang, C.; Jin, H.; Caligiuri, V.; Marras, S.; Schirato, A.; Bursi, L.; Alabastri, A.; Weng, S.; Douaki, A.; Lanzavecchia, G.; Marri, I.; Krahne, R.; Maccaferri, N.; Zheng, Z.; Jin, S.; Garoli, D. Layered Nanoporous Platforms for SERS Sensing. *Adv. Mater. Interfaces* **2026**, e70543. https://doi.org/10.1002/admi.70543.

(59) Hubarevich, A.; Huang, J.; Giovannini, G.; Schirato, A.; Zhao, Y.; Maccaferri, N.; De Angelis, F.; Alabastri, A.; Garoli, D. λ-DNA through Porous Materials—Surface-Enhanced Raman Scattering in a Simple Plasmonic Nanopore. *J. Phys. Chem. C* **2020**, *124* (41), 22663–22670. https://doi.org/10.1021/acs.jpcc.0c06165.